\documentclass[reprint, amsmath, amssymb, aps, superscriptaddress]{revtex4-1}
\usepackage{format}
\allowdisplaybreaks

\begin{document}

\preprint{APS/123-QED}

\title{ Anisotropic acoustics in  dipolar Fermi gases }

\author{Reuben R. W. Wang}
\author{John L. Bohn}
\affiliation{JILA, University of Colorado, Boulder, Colorado 80309, USA}

\date{\today} 

\begin{abstract}

We consider plane wave modes in ultracold, but not quantum degenerate, dipolar Fermi gases in the hydrodynamic limit.  Longitudinal waves present anisotropies in both the speed of sound and their damping, and experience a small, undulatory effect in their flow velocity.  Two distinct types of shear waves appear, a ``familiar" one, and another that is accompanied by nontrivial density and temperature modulations.  We propose these shear modes as an experimental means to measure the viscosity coefficients, including their anisotropies.

\end{abstract}

\maketitle

\section{ \label{sec:introduction} Introduction }

Recent experiments have achieved the trapping and cooling of magnetic lanthanide atoms \cite{Griesmaier05_PRL, Lu11_PRL, Aikawa12_PRL, Tang15_PRA, Phelps20_arxiv, Patscheider21_PRA, Chomaz22_IOP}, and heteronuclear polar molecules \cite{Sage05_PRL, Ni08_Sci, Anderegg17_PRL, Voges20_PRL, Valtolina20_Nat} to ultracold temperatures, ushering in a new era of dipolar physics.
In particular, the realization of collisional shielding in highly polar molecules with electric fields \cite{Valtolina20_Nat, Li21_Nat} and microwaves \cite{Anderegg21_Sci, Schindewolf22_Nat}, now permit investigations of collective dynamics influenced by both long-range and collisional interactions.  
These two effects only become comparable at large densities and dipole moments, establishing the gas as hydrodynamic and therefore more appropriately described as a fluid.    

Classical fluids are known to host a vast variety of dynamical phenomena, attributed to the strong nonlinearities in the governing equations of motion \cite{Landau13_Els}. But even before the onset of nonlinear flows, linear hydrodynamics can already be fascinating \cite{Drazin04_CUP}, made evermore so with dipolar interactions.  In polar molecular gases, tunability of the dipole moment \cite{Karman18_PRL, Karman20_PRA, Lassabliere22_PRA} and its orientation provide a handle to control the influence of dipolar character on collective gas behavior \cite{Quemener12_CR, Wang22_PRA, Wang22_PRA2}. 
For instance, we find that a thermal gas with comparable dipolar mean-field and kinetic energies hosts an anisotropic speed of sound. Unlike its degenerate counterpart, however, nondegenerate gases are prone to the incoherent process of dipolar collisions which impede sound propagation. Of interest here are gases of fermionic molecules where dipolar scattering is universal \cite{Bohn14_PRA}. The thermalizing effect of these collisions manifests in a continuum theory as thermal conductivity and viscosity, quantities that require a tensorial formulation by virtue of the scattering anisotropy. From a microscopic description, we derive these so-called transport tensors for fermions to first order in the Chapman-Enskog fashion \cite{Chapman90_CUP}.  

Although attributed to equilibration, viscosity can in fact be utilize to take the gas out of equilibrium by means of shear flows. When laminar, these flows decay away from the source with characteristic penetration depths that are directly related to the viscosity coefficients. We present these relations explicitly, which could allow for a measurement of the viscosity coefficients from shear flow experiments. 

The remainder of the paper is organized as follows: In Sec.~\ref{sec:linear_hydrodynamics}, we present the general fluid equations of motion for nondegenerate dipolar Fermi gases and linearize it. Propagating modes of the fluid are studied in Sec.~\ref{sec:dipolar_acoustics}, from which a universal anisotropic speed of sound is obtained. A class of mode solutions only present in thermoviscous fluids is studied in Sec.~\ref{sec:thermoviscous_modes}, following which concluding remarks are drawn in Sec.~\ref{sec:conclusions}.

\section{ Linear Dipolar Hydrodynamics \label{sec:linear_hydrodynamics} }

A gas is said to be hydrodynamic when collisions result in fast local thermalization, as occurs when the molecular mean free path is much smaller than a characteristic physical length of the system. This ratio of length scales is often referred to as the Knudsen number, Kn \cite{Chapman90_CUP}.
When hydrodynamic, gases are best described in terms of the continuous field variables of density $\rho$, flow velocity $\boldsymbol{U}$ and kinetic temperature $T$ \cite{Fetter03_Dover}, each of which are dependent on space and time. These variables undergo dynamics governed by the continuity \cite{Reynolds1903_UP}, Navier-Stokes \cite{Navier23_MASIF, Stokes07_SEG} and temperature balance equations \cite{deGroot13_DP}
\begin{subequations} \label{eq:continuum_conservation_laws}
\begin{align}
    \frac{ \partial \rho }{ \partial t } + \partial_j \left( { \rho U_j } \right) 
    &=
    0, \\
    \frac{ \partial \left( \rho U_i \right) }{ \partial t } + \partial_j \left( \rho U_j U_i \right)
    &=
    \partial_j \tau_{i j} - \partial_i P - \frac{ \rho }{ m } \partial_i V(\boldsymbol{r}), \\
    \frac{ \partial (\rho T) }{ \partial t } + \partial_j \left( \rho T U_j \right) 
    &=
    \frac{ 2 m }{ 3 k_B } \big[ ( \tau_{i j} - P \delta_{ij} ) \partial_j U_i \\
    &\quad\quad\quad\quad\quad\: - \kappa_{i j} \partial_i \partial_j T \big], \nonumber 
\end{align}
\end{subequations}
where $P = n k_B T$ is the thermodynamic pressure, $\kappa_{ij}$ is the rank-2 thermal conductivity tensor and $\tau_{i j} = \mu_{i j k \ell} \partial_{\ell} U_k$ is the viscous stress tensor with rank-4 viscosity tensor $\mu_{i j k \ell}$. Repeated indices are assumed to be summed over. 
The external potential in consideration here is that from the dipolar mean-field (DMF) $V(\boldsymbol{r}) = n(\boldsymbol{r}) * \Phi_{\rm dd}\big( \boldsymbol{r} \big)$ \footnote{
One might be concerned that the DMF is already treated microscopically when deriving the transport tensors, and so will be double counted if also included in Eq.~(\ref{eq:continuum_conservation_laws}). However, we show in App.~\ref{app:Boltzmann_equation} that this is not the case since the DMF vanishes in the first order Chapman-Enskog expansion. 
}, 
where $*$ denotes a convolution and
\begin{align} \label{eq:dipole_potential}
    \Phi_{\text{dd}}(\boldsymbol{r}) 
    &= 
    \frac{ d^2 }{ 4 \pi \epsilon_0 }
    \frac{ 1 - 3 ( \hat{\boldsymbol{r}}^T \hat{\boldsymbol{{\cal E}}} )^2 }{ r^3 },
\end{align}
is the dipole-dipole interaction potential between 2 point electric dipoles of dipole moment $d$, aligned along the dipole axis $\hat{\boldsymbol{{\cal E}}}$. In the expression above, $\epsilon_0$ is the electric constant. 

\subsection{Linearization}

In this paper, we limit ourselves to linear modes of an initially uniform gas. To this end the dynamical fields are written as
\begin{subequations} \label{eq:fluctuation_variables}
\begin{align}
    \rho( \boldsymbol{r}, t ) 
    &= 
    \rho_0 [ 1 + \chi( \boldsymbol{r}, t ) ], \\
    U_i( \boldsymbol{r}, t ) 
    &= 
    c \xi_i( \boldsymbol{r}, t ), \\
    T( \boldsymbol{r}, t ) 
    &= 
    T_0 [ 1 + \epsilon( \boldsymbol{r}, t ) ], 
\end{align}
\end{subequations}
only varying in space and time via the unit-free fluctuation fields $\chi, \xi_i, \epsilon \ll 1$, with $c = \sqrt{ 5 k_B T_0 / (3 m) }$ being the ideal gas thermal speed of sound. 
To linear order in the fluctuation field variables, the equations of fluid mechanics become
\begin{subequations} \label{eq:linear_conservation_equations}
\begin{align}
    \frac{ \partial \chi }{ \partial t } + c \partial_j \xi_j 
    &\approx 
    0, \\
    \frac{ \partial \xi_i }{ \partial t } + \frac{ 3 }{ 5 } c \partial_i ( \epsilon + \chi ) 
    &\approx 
    \frac{ \mu_{i j k \ell} }{ \rho_0 } \partial_j \partial_{\ell} \xi_k \\
    &\quad
    -
    \frac{ \rho_0 }{ m^2 c } \partial_i 
    \left[ \chi(\boldsymbol{r}) * \Phi_{\rm dd}( \boldsymbol{r} ) \right],
    \nonumber\\
    \frac{\partial \epsilon}{\partial t} + \frac{ 2 }{ 3 } c \partial_j \xi_j 
    &\approx
    \frac{ 2 }{ 3 n_0 k_B } \kappa_{i j} \partial_i \partial_j \epsilon.
\end{align}
\end{subequations} 
Acoustic mode solutions for such a linear dynamical system can be found by employing a plane wave ansatz $\chi, \xi_i, \epsilon \sim \exp[ i ( \boldsymbol{K}^T \boldsymbol{r} - \omega t ) ]$, with which we notice that the DMF potential just becomes a Fourier transform of $\Phi_{\rm dd}(\boldsymbol{r})$:
\begin{align}
    \chi(\boldsymbol{r}) * \Phi_{\rm dd}( \boldsymbol{r} )
    &= 
    \int d^3 r' \chi(\boldsymbol{r} - \boldsymbol{r}^{\prime}) \Phi_{\rm dd}( \boldsymbol{r}^{\prime} ) \nonumber\\
    &= 
    {\chi} e^{i (\boldsymbol{K}^T \boldsymbol{r} - \omega t) } \int d^3 r' e^{ -i \boldsymbol{K}^T \boldsymbol{r}^{\prime} } \Phi_{\rm dd}( \boldsymbol{r}^{\prime} ) \nonumber\\
    &\equiv 
    {\chi} e^{i (\boldsymbol{K}^T \boldsymbol{r} - \omega t) } 
    \varphi_{\rm dd}(\hat{\boldsymbol{K}}, \hat{\boldsymbol{{\cal E}}}),
\end{align}
with $\varphi_{\rm dd}$ already computed in Ref.~\cite{Xiong09_PRA}. 
Computing dispersion relations for the plane wave modes requires knowledge of the thermal conductivity and viscosity tensors. Dipolar collisions, however, necessitate that these objects maintain a construction in their general tensorial forms since the collisional anisotropy prevents the usual reduction to isotropic coefficients \cite{deGroot13_DP}.  As with dipolar bosons \cite{Wang22_PRA, Wang22_PRA2}, analytic expressions for the these tensors can be obtained using the first-order Chapman-Enskog method \cite{Chapman90_CUP, Reif09_Waveland} and utilizing the differential cross section for dipolar fermions found in Ref.~\cite{Bohn14_PRA}. Details of this derivation are provided in appendices~\ref{app:transport_tensors} and \ref{app:Boltzmann_equation}, along with a list of the explicit functional forms for each transport tensor element. 

\subsection{Plane Waves}

Employing a plane wave {\it ansatz} renders derivatives of fluctuations variables $\partial_j \rightarrow i K_j$ and $\frac{ \partial }{ \partial t } \rightarrow -i \omega$, so the differential equations of Eq.~(\ref{eq:linear_conservation_equations}) reduce to the eigensystem
\begin{align}  \label{eq:eigensystem}
    \omega \begin{pmatrix}
        \chi \\
        \boldsymbol{\xi} \\
        \epsilon
    \end{pmatrix}
    =
    \begin{pmatrix}
        0 & c \boldsymbol{K}^T & 0 \\
        \left( \frac{3}{5} c + \frac{ \rho_0 \varphi_{\rm dd} }{ m^2 c } \right) \boldsymbol{K} & 
        i \boldsymbol{\Lambda} & 
        \frac{3}{5} c \boldsymbol{K} \\
        0 & \frac{2}{3} c \boldsymbol{K}^T & i \Gamma 
    \end{pmatrix}
    \begin{pmatrix}
        \chi \\
        \boldsymbol{\xi} \\
        \epsilon
    \end{pmatrix},
\end{align}
having defined the thermal conductivity and viscosity associated rates
\begin{subequations}
\begin{align}
    \Gamma(\Theta) 
    &= 
    -\frac{ 2 }{ 3 n_0 k_B } \kappa_{ij} K_i K_j, \\
    \Lambda_{i k}(\Theta) 
    &= 
    -\frac{ 1 }{ \rho_0 } \mu_{i j k \ell} K_j K_{\ell},
\end{align}
\end{subequations} 
respectively. We have defined $n_0 = \rho_0 / m$ as the equilibrium number density.
With no other processes to break the symmetry of the system, the anisotropy that arises in the mode solutions only depends on the relative angle between the dipole orientation $\hat{\boldsymbol{{\cal E}}}$ and plane wave propagation direction $\hat{\boldsymbol{K}}$. Thus, all essential physics is captured by setting $\hat{\boldsymbol{K}} = \hat{\boldsymbol{z}}$ 
but allowing $\Theta$ to vary. In these coordinates, which we assert for the remainder of this paper, the transport associated rate functions have the forms (See App.~\ref{app:transport_tensors})
\begin{widetext}
\begin{subequations} \label{eq:transport_rates}
\begin{align}
    \Gamma(\Theta) 
    &= 
    \frac{ 875 K^2 }{ 12288 a_d^2 n_0 \sqrt{ \pi m \beta_0 } }
    \left( \cos (2 \Theta ) - 5 \right), \\
    \boldsymbol{\Lambda}(\Theta) 
    &= 
    \frac{ 5 K^2 }{ 2048 a_d^2 n_0 \sqrt{ \pi m \beta_0 } }
    \left(
    \begin{array}{ccc}
     \frac{ 9 }{ 4 } (13 \cos (4 \Theta ) - 29) & 0 & \frac{ 3 }{ 4 } ( 14 \sin (2 \Theta ) - 39 \sin (4 \Theta ) ) \\
     0 & 9 (5 \cos (2 \Theta ) - 9) & 0 \\
     \frac{ 3 }{ 4 } ( 14 \sin (2 \Theta ) - 39 \sin (4 \Theta ) ) & 0 & \frac{ 1 }{ 4 } ( 84 \cos (2 \Theta ) - 117 \cos (4 \Theta ) - 415 ) \\
    \end{array}
    \right),
\end{align}
\end{subequations}
\end{widetext}
where $\beta_0 = (k_B T_0)^{-1}$ is the usual inverse temperature, while $\varphi_{\rm dd}$ reduces to the simple form
\begin{align}
    \varphi_{\rm dd}(\Theta)
    &=
    \frac{ d^2 }{ 3 \epsilon_0 } (3 \cos^2 \Theta - 1). 
\end{align}

\subsection{ Mode Frequencies }

Any fluid dynamics resultant from Eq.~(\ref{eq:eigensystem}) is fully described by normal mode solutions, comprising of mode frequencies $\omega_a$, and their corresponding mode amplitudes $\boldsymbol{\psi}_a = ( \chi_a, \boldsymbol{\xi}_a, \epsilon_a )$ with $a = 1$ to $5$. 
To obtain these normal modes analytically, we consider long wavelength excitations such that $\delta = K L \ll 1$, where $L = ( \overline{\sigma} n_0 )^{-1}$ is the molecular mean free path and $\overline{\sigma} = 32 \pi a_d^2 / 15$ is the angular averaged total cross section \cite{Bohn14_PRA} with dipole length $a_d = d^2 m / ( 8 \pi \epsilon_0 \hbar^2 )$. 
Since $\mu_{i j k \ell}, \kappa_{i j} \sim \delta$, long wavelengths then permit series expansions of the mode solutions in increasing powers of the transport tensors via Taylor expansions in $\delta$.   
By diagonalizing Eq.~(\ref{eq:eigensystem}) (refer to App.~\ref{app:mode_derivation}), we obtain the following mode frequency solutions 
\begin{subequations} \label{eq:firstorder_omega_solutions}
\begin{align} 
    \omega_{\pm }
    &=
    \pm K \sqrt{ c^2 + \mathscr{E}_{\rm dd}(\Theta) }
    +
    \frac{ i }{ 2 } \left( \Lambda_{33} + \frac{ 2 \Gamma c^2 }{ 5 ( c^2 + \mathscr{E}_{\rm dd} ) } \right), \\
    \omega_{\mu, 1} 
    &= 
    i \Lambda_{11}, \\
    \omega_{\mu, 2} 
    &= 
    i \Lambda_{22}, \\
    \omega_{\kappa} 
    &= 
    i \Gamma \frac{ \left( 3 c^2 + 5 \mathscr{E}_{\rm dd} \right) }{ 5 \left( c^2 + \mathscr{E}_{\rm dd} \right)},
\end{align}
\end{subequations}
to first order in $\delta$, where  
$\mathscr{E}_{\rm dd}(\Theta) = n_0 \varphi_{\rm dd}(\Theta) / m$
is the specific DMF energy. The corresponding modes and their interpretation will be given in the following sections. 

\section{ Anisotropic Sound \label{sec:dipolar_acoustics} }

\subsection{Sound Velocity}

The first two modes, with frequencies $\omega_{\pm}$, represent propagating, longitudinal sound waves.
Taking the long-wavelength limit identifies the anisotropic speed of sound $c_{\rm dd}(\Theta) = \sqrt{ c^2 + \mathscr{E}_{\rm dd}(\Theta) }$.
This is usefully written in terms of the dimensionless parameter
\begin{align}
    \eta 
    =
    \frac{ 2 \pi \rho_0 }{ 5 }
    \left( \frac{ d^2 }{ \epsilon_0 } \right)
    \left( \frac{ \lambda_{\rm th} }{ h } \right)^2,
\end{align}
which is a function of the thermal de Broglie wavelength $\lambda_{\rm th} = h / \sqrt{ 2 \pi m k_B T_0 }$ with $h$ as Planck's constant. The speed of sound is given in terms of $\eta$ as
\begin{align} \label{eq:speed_of_sound_eta}
    c_{\rm dd}(\Theta) 
    &=
    c \sqrt{ 1 + \eta \left( { 3 \cos^2 \Theta - 1 } \right) },
\end{align}
which is completely independent of the quantum statistics of the constituent molecules.
As written, the speed of sound in a normal dipolar gas has an anisotropy similar to that for a dipolar Bose-Einstein condensate (DBEC) \cite{Santos00_PRL, Bismut12_PRL, Lima12_PRA}, but with temperature replacing the role of quantum fluctuations.
The quantity $\eta$ compares the magnitude of the DMF with thermal energies, which at a fixed temperature $T_0$, varies by means of the background density $\rho_0$ and dipole moment $d$ \cite{Anderegg21_Sci, Li21_Nat, Schindewolf22_Nat, Lassabliere22_PRA}.

For propagation of sound waves along the direction of dipole polarization, $\Theta \sim 0$, this propagation is stable, that is, the value of $c_{\rm dd}$  remains real-valued.  However, for sufficiently large $\eta$, the speed of sound develops a significant anisotropy due to the growing contribution from the DMF. 
Going past the critical value of $\eta_c(\Theta) \equiv -( 3 \cos^2\Theta - 1 )^{-1}$, DMF interactions may overcome the thermal kinetic energy, causing $c_{\rm dd}(\theta_K)$ to become imaginary \footnote{ 
Even at this higher density of $n_0 = 2 \times 10^{12}$ cm$^{-3}$, the transport coefficients presented in Eqs.~(\ref{eq:thermal_conductivities}) and (\ref{eq:viscosities}) are accurate to 10\% in higher density corrections \cite{Chapman90_CUP}.
}.
An imaginary speed of sound indicates a dipolar instability, also predicted and observed in DBEC \cite{Santos00_PRL, Lahaye08_PRL, Baranov08_PR, Bohn09_LP, Lahaye09_IOP, Norcia21_NatPhys}. 
Notably, $\eta_c$ is only well defined within the interval bounded by the dipolar magic angles $\Theta_{\rm magic} \approx 54.7^{\circ}$ and $125.3^{\circ}$, at which $\varphi_{\rm dd}(\Theta_{\rm magic}) = 0$ 
\footnote{
The critical value $\eta_c$ appears to be a pole of certain solutions in Eqs.~(\ref{eq:firstorder_omega_solutions}). This divergence is however, only a feature of the series expansion in $\delta$. Exact solutions to Eq.~(\ref{eq:eigensystem}) remain finite valued across $\eta = \eta_c$. 
}. 
Within the range of dipolar magic angles, $\eta_c$ has a minima of $1$ at $\Theta = 90^{\circ}$.

\subsection{Undulating Sound Waves}

\begin{figure}[ht]
    \centering
    \includegraphics[width=\columnwidth]{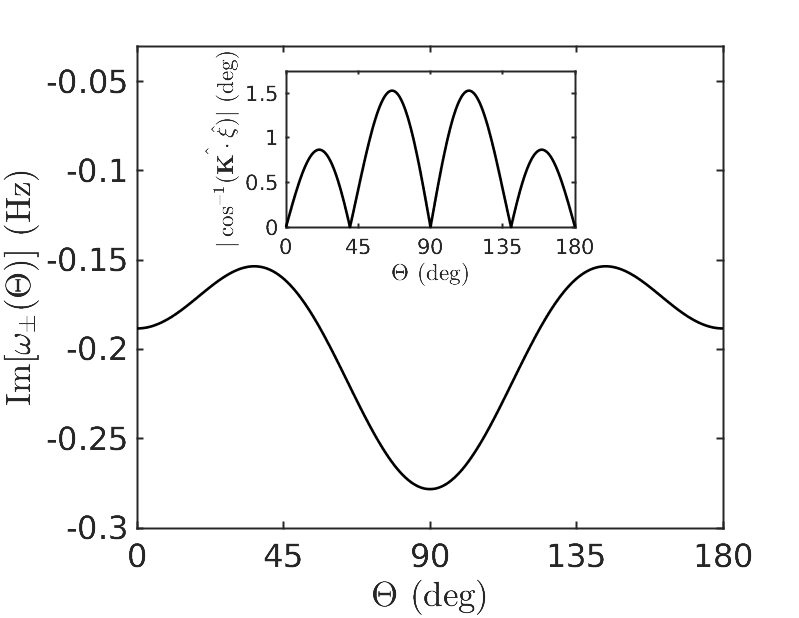}
    \caption{ Imaginary part of the propagating mode frequency solutions ${\rm Im}[ \omega_{\pm}(\Theta) ]$ in Hertz (Hz), as a function of the dipole tilt angle $\Theta$ in degrees (deg) at $\eta \approx 0.04$. The figure inset plots the absolute relative angle between $\hat{\boldsymbol{K}}$ and $\hat{\boldsymbol{\xi}}$, $| \cos^{-1}( \hat{\boldsymbol{K}} \cdot \hat{\boldsymbol{\xi}} ) |$, also as a function of $\Theta$. }
    \label{fig:sound_mode_anisotropy}
\end{figure}

The physics of sound gets more interesting at finite $K$, where transport tensors now enter the dynamical arena.  
To ground our discussions, we envision an experiment with a box-trapped \cite{Navon21_Nat} uniform density sample of microwave shielded $^{23}$Na$^{40}$K molecules cooled to $T_0 = 250$ nK \cite{Schindewolf22_Nat}.
When $\eta \ll 1$, say at $n_0 = 10^{12}$ cm$^{-3}$ and $d = 0.75$ D ($\eta \approx 0.04$), the DMF effects and therefore $\mathscr{E}_{\rm dd}$, becomes negligible compared to kinetic processes. In this regime, we see that the speed of sound reverts to that of an ideal gas $c_{\rm dd} \approx c$, while the imaginary part of $\omega_{\pm}$ is strictly negative, leading to sound attenuation. Resulting directly from dipolar collisions, the observed attenuation for a long wavelength excitation of $\delta = 0.1$ is anisotropic, varying by a factor of $\sim 2$ with $\Theta$ as shown in Fig.~\ref{fig:sound_mode_anisotropy}.  

The $\omega_{\pm}$ modes have associated eigenvectors of the form
\begin{align} \label{eq:propagating_modes}
    \boldsymbol{\psi}_{\pm} 
    &=
    \begin{pmatrix}
        1  \\
        0 \\
        0 \\
        \pm \frac{ c_{\rm dd} }{ c } \\
        \frac{2}{3}
    \end{pmatrix} 
    +
    \begin{pmatrix}
        \mp i \frac{ \Gamma }{ c_{\rm dd} K }  \\
        i \frac{ \Lambda_{13} }{ c K } \\
        0 \\
        - i \left( \frac{ \Gamma }{ 5 c K }  \left( 4 + \frac{ \mathscr{E}_{\rm dd} }{ c_{\rm dd}^2 } \right) + \frac{ \Lambda_{33} }{ 2 c K } \right) \\
        0
    \end{pmatrix}, 
\end{align}
defined up to an overall scale factor.  The first term of the sum in the expression above has nonzero and comparable amplitudes in the fractional density shift $\chi$, fractional $z$-velocity $\xi_z$, and fractional temperature shift $\epsilon$, as expected for a longitudinal wave propagating in the $z$ direction.
The second term in $\psi_{\pm}$ shows the additional effects introduced by viscous and thermal damping.  Specifically, terms in the density and $z$-velocity are explicitly damped, while the temperature is not yet damped at this level of approximation.  Along with these effects, a new one appears, namely, a damped motion in the $x$-velocity $\xi_x$. 

These eigenvectors therefore indicate that, despite initiating sound along $\hat{\boldsymbol{K}} = \hat{\boldsymbol{z}}$, the fluctuations in flow velocity could occur in a slightly different direction depending on the dipole orientation. 
That is, in these plane wave solutions, the fluid flow along $z$ alternately compresses and rarefies the gas, while the fluid velocity simultaneously alternates between flow along the $+x$ and $-x$ directions. The general fluid motion is therefore of a slightly undulatory nature.  
This effect, albeit small, is one unique to anisotropic transport which we illustrate with a plot of the absolute relative angle between $\hat{\boldsymbol{K}}$ and $\hat{\boldsymbol{\xi}}$, $| \cos^{-1}( \hat{\boldsymbol{K}} \cdot \hat{\boldsymbol{\xi}} ) |$, against $\Theta$ in the inset of Fig.~\ref{fig:sound_mode_anisotropy}. These weak transverse motions are potentially observable in Doppler spectroscopy of the undulating molecules.

\section{ Thermoviscous Modes \label{sec:thermoviscous_modes} }

\begin{figure}[ht]
    \centering
    \includegraphics[width=\columnwidth]{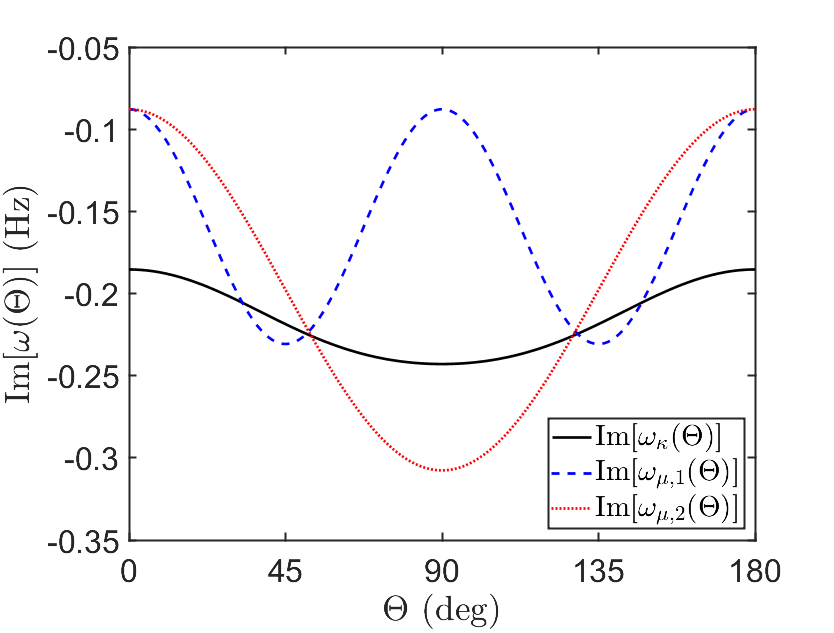}
    \caption{ The imaginary parts of the mode frequency solutions, $\Im[\omega_{\kappa}]$ (solid black curve), $\Im[\omega_{\mu,1}]$ (dashed blue curve) and $\Im[\omega_{\mu,2}]$ (dotted red curve), as a function of $\Theta$ in a gas of $^{23}$Na$^{40}$K molecules with $n_0 = 10^{12}$ cm$^{-3}$ and $d = 0.75$ D ($\eta \approx 0.04$).  }
    \label{fig:anisotropic_damping_lowdensity}
\end{figure}

Even in the absence of sound, a silent hydrodynamic gas of dipoles still has a story to tell. 
This narrative is populated by the latter 3 modes of Eq.~(\ref{eq:firstorder_omega_solutions}), all of which have purely imaginary frequencies for any value of $\eta$ and $\Theta$.  
Once again with $\eta \approx 0.04$ and $\delta = 0.1$, which suppresses effects from the DMF, anisotropic damping is accentuated in these silent modes with plots of the imaginary parts of their mode frequencies in Fig.~\ref{fig:anisotropic_damping_lowdensity}.

One of these modes, with frequency $\omega_{\mu,2}$, has a particularly simple form:
\begin{align} \label{eq:y_shear_mode}
    \boldsymbol{\psi}_{\mu, 2}
    = 
    \begin{pmatrix}
        0 \\
        0 \\
        1 \\
        0 \\
        0
    \end{pmatrix}.
\end{align}
This mode consists exclusively of flow velocity in the $\pm y$ directions, the velocity being sinusoidally modulated along $z$ with wavelength $2 \pi / K$. If one were to ``grab'' the $z=0$ layer of the fluid and shake it with frequency $|\omega_{\mu,2}|$, a shear wave would thus develop.  This is an overdamped mode, hence its amplitude reaches only to approximately a certain penetration depth, defined as the inverse absolute imaginary part of the wave-number \cite{Fetter03_Dover, Wang22_PRA2}:
\begin{align}
    r_{\mu,2} 
    &=
    \sqrt{ \frac{ \mu_{2323} }{ 2 \omega \rho_0 } }
    \approx
    \frac{ 3 }{ 64 }
    \sqrt{ \frac{ 5 ( 9 - 5 \cos (4 \Theta ) ) }{ \omega a_d^2 n_0 \sqrt{\pi m \beta_0} } }. 
\end{align}
The expression above is obtained from Eq.~(\ref{eq:firstorder_omega_solutions}), by instead solving for $K$ in terms of $\omega$.
Such waves are, of course, already familiar in ordinary, isotropic fluids.

A shear mode with fluid flow in the $\pm x$ direction is, however, affected by the anisotropy of the scattering (recall that the dipolar orientation lies in the $x,z$ plane). The $x$-shear mode is given by
\begin{align} \label{eq:x_shear_mode}
    \boldsymbol{\psi}_{\mu, 1}
    &= 
    \begin{pmatrix}
        0 \\
        \left( 1 - \frac{ \Gamma }{ \Lambda_{11} } \right) \frac{ 5 c_{\rm dd}^2 }{ 2 c^2 } 
        +
        \frac{ \Gamma }{ \Lambda_{11} } \\
        0 \\
        0 \\
        0
    \end{pmatrix} 
    +
    \begin{pmatrix}
        i \frac{ \Gamma }{ c K } \frac{ c^2 \Lambda_{13} }{ c^2_{\rm dd} { \Lambda_{11} } } \\
        0 \\
        0 \\
        0 \\
        -i \frac{ 5 \Lambda_{13} }{ 3 c K }
    \end{pmatrix}.
\end{align}
Here again, the first term accounts for the dominant motion, namely, oscillations in the $\pm x$ directions induced by shear. 
The penetration depth for this $x$-shear mode under an oscillatory shear drive is, not surprisingly, also dipole angle dependent, and given by the relation
\begin{align}
    r_{\mu,1} 
    &=
    \sqrt{ \frac{ \mu_{1313} }{ 2 \omega \rho_0 } }
    \approx
    \frac{ 3 }{ 128 }
    \sqrt{ \frac{ 5 ( 29 - 13 \cos (4 \Theta ) ) }{ \omega a_d^2 n_0 \sqrt{\pi m \beta_0} } }.
\end{align}

The second term in $\psi_{\mu.1}$ denotes additional, accompanying effects associated with this shear mode, in this case damped modulations in the density and temperature fields. In this circumstance, where the dipoles are oriented somewhere in the $x,z$ plane, and the shear flow in the $x$ direction, the anisotropy of the collision cross section is capable of shoveling both matter and kinetic energy preferentially into the $\pm z$ directions, the same as it ordinarily does for momentum.

Because shear modes are a direct consequence of viscosity in the gas, they present themselves as an experimental means to measure the viscosity coefficients \footnote{
We point out that shear viscosity has also been measured in harmonically confined unitary Fermi gases with breathing mode damping experiments \cite{Cao11_Sci}. These measurement were, however, performed for non-dipolar gases with a single shear viscosity coefficient. 
}. We propose an oscillatory shear layer of constant frequency $\omega$, realizable in ultracold experiments with a time-dependent box trap boundary condition. This oscillation would induce a shear excitation that travels orthogonal to the shear layer with the $\Theta$-dependent penetration depths provided above.
Seeing that the penetration depths scale as $\sim \sqrt{ \mu_{i j k \ell} } \sim \sqrt{ L }$, ensuring $r_{\mu} \ll L$ so as to remain hydrodynamic, can be enforced by tuning the dipole moment appropriately.  
We therefore consider $d = 2.5$D at $n_0 = 10^{12}$cm$^{-3}$ ($\eta \approx 0.46$), with an oscillation frequency of $\omega = 2\pi 100$ Hz. These parameters result in $r_{\mu} / L \approx 18$ to $35$, with values depending on the dipole orientation $\Theta$.

\begin{figure}[ht]
    \centering
    \includegraphics[width=\columnwidth]{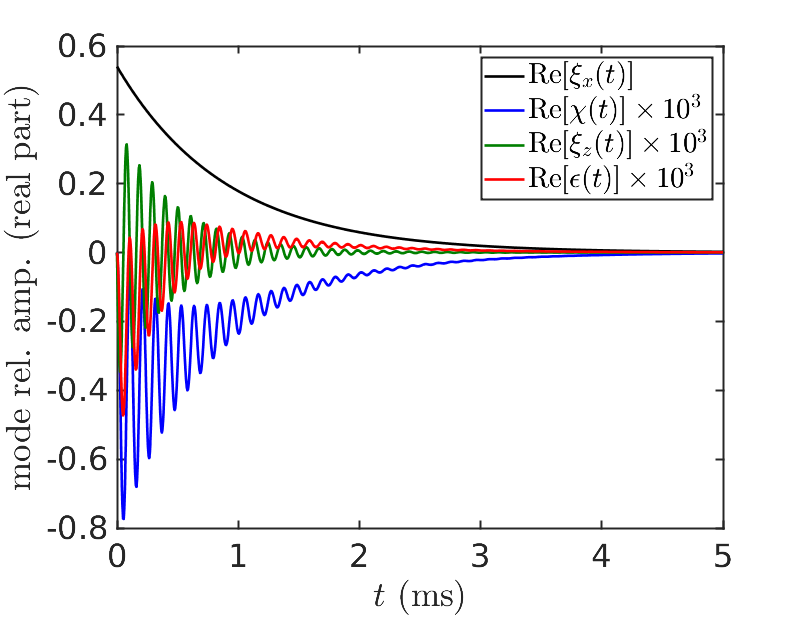}
    \caption{ Relative fluid variable fluctuation amplitudes ${\xi}_x(t)$ (black curve), $\chi(t)$ (blue curve), ${\xi}_z(t)$ (green curve), $\epsilon(t)$ (red curve) as a function of time $t$ at $z/L = 10$, after an impulse shear flow perturbation along $\hat{\boldsymbol{x}}$. The dipoles are oriented with $\Theta = \pi/4$. The relative amplitudes for $\chi(t), \epsilon(t)$ and $\xi_z(t)$ are rescaled by a factor of $10^3$ for clarity of presentation. }
    \label{fig:xshear_time_evo}
\end{figure}

To demonstrate this anomalous shear excitation, we plot the time evolution of relative mode amplitudes $\boldsymbol{\psi}(t)$ in Fig.~\ref{fig:xshear_time_evo}, after an impulse shear flow perturbation along $\hat{\boldsymbol{x}}$ with $\Theta = \pi/4$. We find that along with $\chi$ and $\epsilon$, $\xi_z$ is also subsequently excited from this perturbation. 
In the plot, the relative amplitudes for $\chi(t), \epsilon(t)$ and $\xi_z(t)$ are rescaled by a factor of $10^3$ for ease of visualization, but indicate that this effect is indeed a small one. 
Temperature field variations via the introduction of heat into a fluid are commonly referred to as entropy waves \cite{Chu58_JFM, Morgans16_IJSCD}, which if initiated by laminar shear flow, motivates the title ``shear-entropy waves". This phenomena is, to our knowledge, not present in existing shear flow literature.

Not forgetting the fifth and final mode $\boldsymbol{\psi}_{\kappa}$, we find that
in the range of $\Theta$ with $\mathscr{E}_{\rm dd} < 0$, $\omega_{\kappa}$ can vanish identically at suitably large values of $n_0$ and $a_d$ such that $\mathscr{E}_{\rm dd} = -3 c^2 / 5$ (i.e. $\eta = 3 \eta_c / 5$). Satisfying this condition would lead to a mode where $\epsilon = \xi_x = \xi_y = \xi_z = 0$ but $\chi \neq 0$, as made evident by its functional form 
\begin{align} \label{eq:density_mode}
    \boldsymbol{\psi}_{\kappa}
    &= 
    \begin{pmatrix}
        0 \\
        \frac{3}{5} \left( 1 - \frac{ \Gamma }{ \Lambda_{11} } \frac{ 3 c^2 + 5 \mathscr{E}_{\rm dd} }{ 5 c_{\rm dd}^2 } \right) \left( 1 + \frac{ 5 \mathscr{E}_{\rm dd} }{ 3 c^2 } \right) \\
        0 \\
        0 \\
        0
    \end{pmatrix} \nonumber\\
    &\quad +
    \begin{pmatrix}
        i \frac{ \Lambda_{13} }{ c K } \\
        0 \\
        0 \\
        0 \\
        -i \frac{ \Lambda_{13} }{ c K } \left( 1 + \frac{ 5 \mathscr{E}_{\rm dd} }{ 3 c^2 } \right)
    \end{pmatrix}.
\end{align}
Such a mode implies the existence of long-lived density modulations due to a balance between thermal and dipolar mean-field energies. We leave further analysis of this phenomenon to a future work.

\section{ Conclusions \label{sec:conclusions} }

We have formulated a comprehensive theory of normal dipolar Fermi gases in the hydrodynamic regime, including effects from both the DMF and collisions. When comparable to the thermal energy, the DMF introduces a significant anisotropy to the speed of sound. 
That is, sound travels faster when parallel to the dipole orientation but slower when orthogonal. 
If too large, the DMF can completely overpower thermal molecular motion, triggering dynamical instability (i.e. the speed of sound becomes imaginary).  
As a consequence of the sound speed dipole angle dependence, these instabilities only occur in waves that propagate at angles within $\Theta \approx 54.7^{\circ}$ and $125.3^{\circ}$, relative to the dipole orientation. Outside this angular range, linear wave excitations remain completely robust against dipolar collapse.

Collisions on the other hand, serve to return the gas to hydrodynamic equilibrium, the route of which is also anisotropic. As in the case of bosons \cite{Wang22_PRA, Wang22_PRA2}, thermal conductivity and viscosity encompass these local thermalization effects and result in the anisotropic damping of sounds waves. Additionally, we find that these transport tensors lead to a minute undulatory divergence between the wave propagation $\hat{\boldsymbol{K}}$ and fluid flow $\hat{\boldsymbol{\xi}}$ directions, an effect not possible in dipolar superfluids.   
Not surprisingly, anisotropy is also present in shear excitations that are directly consequent of viscous stresses. We therefore suggest that experimental realizations of shear waves could permit measurements of the viscosity coefficients. A curiosity of laminar shearing a dipolar gas, is that the viscid flow could incite an anomalous density and temperature excitation we identify as shear-entropy waves. This effect was one we found no analogy to in the literature.

With dense long-lived samples of polar molecules now accessible to the ultracold community, we expect the phenomena presented in this work and more yet to be explored with our theory, are experimentally achievable with current technologies.
A direction for future work could therefore be to solve the full nonlinear fluid equations, where we expect to find a rich tapestry of hydrodynamic phenomena.
Additionally, recent experiments of electric field shielded KRb molecules have suggested suppression of both two and three-body losses \cite{Matsuda20_Sci, Li21_Nat, Schmidt22_PRR}. This opens opportunities for hydrodynamic studies after the onset of dipolar collapse \cite{Bohn09_LP}, where sustaining a large fraction of molecules past the instability could permit turbulent cascades from strong nonlinear flows \cite{Drazin04_CUP}.

\begin{acknowledgments}

This work is supported by the National Science Foundation under Grant Number PHY2110327. 

\end{acknowledgments}

\appendix

\section{ Evaluation of the Transport Tensors \label{app:transport_tensors} }

In a nondegenerate gas, local equilibrium occurs by means of dipolar collisions parameterized by the dipole length $a_d$. 
Close to local thermal equilibrium, re-equilibration processes are encapsulated by the transport tensors of viscosity and thermal conductivity, derivable from a microscopic picture with methods established by Chapman and Enskog \cite{Chapman90_CUP}. 
Within length scales on the order of the molecular mean-free path, molecular interactions are dominated by collisional processes. The local distribution of molecules thus has dynamics well described by the Boltzmann transport equation 
\begin{subequations}
\begin{align} \label{eq:Boltzmann_equation}
    & \left( \frac{ \partial }{ \partial t } + v_i \partial_i \right) f(\boldsymbol{r}, \boldsymbol{v}) = {\cal C}[ f(\boldsymbol{r}, \boldsymbol{v}) ], \\
    & {\cal C}[ f ] = \int d\Omega' \frac{d\sigma}{d\Omega'} \int d^3 v_1 \abs{\boldsymbol{v} - \boldsymbol{v}_1} \left( f' f_1' - f f_1 \right),
\end{align}
\end{subequations}
where $f(\boldsymbol{r}, \boldsymbol{v})$ is the phase space distribution function and ${\cal C}[f]$ is the two-body collision integral. All repeated indices are summed over and primes denote post-collision velocities for pairs of colliding molecules with incoming velocities $\boldsymbol{v}$ and $\boldsymbol{v}_1$. We also adopt the compact notation $f_1 = f(\boldsymbol{r}, \boldsymbol{v}_1)$ and $f' = f(\boldsymbol{r}, \boldsymbol{v}')$.   
The gas number density is given by $n(\boldsymbol{r}, t) = \int d^3 v f(\boldsymbol{r}, \boldsymbol{v}, t)$, which is only dependent on temperature at thermal equilibrium, $n_0 = n_0(\beta)$. Thermal equilibrium also imposes a Boltzmann velocity distribution  
\begin{align} \label{eq:equilibrium_ansatz}
    f_0(\boldsymbol{u}, \beta) &= n_0(\beta) c_0(\boldsymbol{u}, \beta) \nonumber\\
    &= n_0(\beta) \left( \frac{ m \beta }{ 2 \pi } \right)^{3/2} \exp\left( - \frac{ m \beta }{ 2 } \boldsymbol{u}^2 \right),
\end{align}
where $\beta = (k_B T)^{-1}$, $\boldsymbol{u}^2 = u_k u_k$, and $\boldsymbol{u}(\boldsymbol{r}) = \boldsymbol{v} - \boldsymbol{U}(\boldsymbol{r})$ is the  
peculiar velocity, defined as the molecular velocity $\boldsymbol{v}$ relative to the flow velocity $\boldsymbol{U}(\boldsymbol{r}, t) = n(\boldsymbol{r}, t)^{-1} \int d^3 v f(\boldsymbol{r}, \boldsymbol{v}, t) \boldsymbol{v}$

Close to thermal equilibrium, the molecular distribution fluctuates as
\begin{align} \label{eq:close2equilibrium_ansatz}
    & f(\boldsymbol{r}, \boldsymbol{u}, \beta) \approx f_0(\boldsymbol{u}, \beta) [ 1 + \Phi(\boldsymbol{r}, \boldsymbol{u}, \beta) ], 
\end{align}
with a perturbation function $\Phi$, that must satisfy 
\begin{subequations} \label{eq:ansatz_conservation_laws}
\begin{align}
    & \int d^3 u f_0(\boldsymbol{u}) \Phi(\boldsymbol{r}, \boldsymbol{u}, \beta) m = 0, \\
    & \int d^3 u f_0(\boldsymbol{u}) \Phi(\boldsymbol{r}, \boldsymbol{u}, \beta) m \boldsymbol{u} = 0, \\
    & \int d^3 u f_0(\boldsymbol{u}) \Phi(\boldsymbol{r}, \boldsymbol{u}, \beta) \frac{1}{2} m \boldsymbol{u}^2 = 0,
\end{align}
\end{subequations}
as a result of mass, momentum and energy conservation respectively. Enskog's prescription of successive approximations then renders the Boltzmann equation, to leading non-trivial order, as
\begin{align} \label{eq:1stOrder_BoltzmannEquation}
    \left( \frac{\partial }{ \partial t } + v_i \partial_i \right) f_0 \approx C[ f_0 \Phi ].
\end{align}
The left-hand side of Eq.~(\ref{eq:1stOrder_BoltzmannEquation}) above evaluates to
\begin{align} \label{eq:Boltzmann_LHS}
    & \left( \frac{\partial }{ \partial t } + v_k \partial_k \right) f_0  
    = f_0 \big[ V_{k} \partial_k ( \ln T ) + m \beta W_{k\ell} D_{k\ell} \big],
\end{align}
as detailed in App.~\ref{app:Boltzmann_equation}, where 
\begin{subequations}
\begin{align}
    & V_i(\boldsymbol{u}) \equiv \left( \frac{ m \beta \boldsymbol{u}^2 }{ 2 } - \frac{ 5 }{ 2 } \right) u_i, \\
    & W_{ij}(\boldsymbol{u}) \equiv u_i u_j - \frac{ 1 }{ 3 } \delta_{ij} \boldsymbol{u}^2, \\
    & D_{ij}(\boldsymbol{U}) \equiv \frac{ 1 }{ 2 } \left( \partial_{j} U_i + \partial_i U_j \right) - \frac{ 1 }{ 3 } \delta_{ij} \partial_k U_k.
\end{align}
\end{subequations}
The collision integral on the right-hand side of Eq.~(\ref{eq:1stOrder_BoltzmannEquation}) is then
\begin{align}
    C[ f ] \approx \int d^3 u_1 & \abs{ \boldsymbol{u} - \boldsymbol{u}_1 } f_0(\boldsymbol{u}) f_0(\boldsymbol{u}_1) \int d\Omega' \frac{ d \sigma }{ d \Omega' } \Delta \Phi, \label{eq:collision_integral}
\end{align}
where $\Delta \Phi = \Phi' + \Phi'_1 - \Phi - \Phi_1$. Since Eq.~(\ref{eq:collision_integral}) is linear in $\Phi$ and Eq.~(\ref{eq:Boltzmann_LHS}) is linear in the quantities $\partial_i \ln T$ and $D_{i j}$, one can infer an {\it ansatz} for the scalar function $\Phi$, of the form 
\begin{align} \label{eq:out_of_equilibrium_ansatz} 
    \Phi(\boldsymbol{u}, \beta) &= {\cal B}_k \partial_k ( \ln T ) + m \beta {\cal A}_{k\ell} D_{k\ell},
\end{align}
where $\boldsymbol{{\cal B}}$ (vector) and $\boldsymbol{{\cal A}}$ (2-rank tensor) are functions of $\boldsymbol{u}$ and $\beta$. Upon comparing terms, $\boldsymbol{{\cal B}}$ and $\boldsymbol{{\cal A}}$ must have the forms
\begin{subequations}
\begin{align}
    & {\cal A}_{ij}(\boldsymbol{u}, n_0, \beta) = W_{k\ell}(\boldsymbol{u}) a_{k \ell i j}(u, n_0, \beta), \\
    & {\cal B}_{i}(\boldsymbol{u}, n_0, \beta) = V_j(\boldsymbol{u}) b_{j i}(u, n_0, \beta),
\end{align}
\end{subequations}
where $u = \abs{\boldsymbol{u}}$, and the coefficients $a_{k \ell m n}(u, n_0, \beta)$ and $b_{k \ell}(u, n_0, \beta)$ are introduced as variational {\it ansatz}. We are thus left with  
\begin{align} \label{eq:Phi_ansatz}
    \Phi(\boldsymbol{u}, \beta) 
    &=
    V_{\ell}(\boldsymbol{u}) b_{\ell k}(n_0, \beta) \partial_k ( \ln T ) \nonumber\\
    &\quad\quad + 
    2 m \beta W_{i j}(\boldsymbol{u}) a_{i j k \ell}(n_0, \beta) D_{k\ell}.
\end{align}

Referring back to Eq.~(\ref{eq:Phi_ansatz}), an average is taken over the molecular distribution by multiplying Eq.~(\ref{eq:1stOrder_BoltzmannEquation}) by $V_i(\boldsymbol{u})$ and $W_{i j}(\boldsymbol{u})$, then integrating over $\boldsymbol{u}$ to give 
\begin{align}
    & \left( \int d^3 u \: f_0( \boldsymbol{u} ) V_i(\boldsymbol{u}) V_j(\boldsymbol{u}) \right) \partial_j ( \ln T ) 
    \\
    &\quad\quad\quad \approx 
    \left( \int d^3 u \: V_i(\boldsymbol{u}) C[ f_0 V_k ] \right) b_{k j} \partial_j ( \ln T ), \nonumber\\
    & 
    \left( \int d^3 u f_0(\boldsymbol{u}) W_{i j} W_{k \ell} \right) D_{ k \ell }
    \\ 
    &\quad\quad\quad \approx 
    \left( -\frac{ \beta }{ n_0 } \int d^3 u \: W_{i j} {\cal C}[ f_0 W_{m n} ] \right) 
    \mu_{ m n k \ell } D_{ k \ell }, \nonumber
\end{align} 
respectively. Evaluating the integrals with the methods in Refs.~\cite{Wang20_PRA, Wang21_PRA} and comparing them to the linear constitutive relations
\begin{subequations}
\begin{align}
    J_i 
    &= 
    -\kappa_{ij} \partial_j T, \\
    \tau_{i j} 
    &=
    2 \mu_{ i j k \ell } D_{ k \ell }, 
\end{align}
\end{subequations}
we obtain the transport tensor of thermal conductivity
\begin{align} \label{eq:thermal_conductivities}
    \boldsymbol{\kappa} 
    =
    \frac{ 175 \mu_0 k_B }{ 8 m }
    \begin{pmatrix}
        5 + \cos(2 \Theta ) & 0 & -\sin(2 \Theta ) \\
        0 & 6 & 0 \\
        -\sin(2 \Theta ) & 0 & 5 - \cos(2 \Theta )
    \end{pmatrix},
\end{align}
and the 13 unique viscosity coefficients
\begin{subequations} \label{eq:viscosities}
\begin{align}
    & \mu_{1111} = \frac{\mu_{0}}{8} (117 \cos (4 \Theta )+84 \cos (2 \Theta )+415), \\
    & \mu_{1113} = -\frac{3 \mu_{0}}{8} (39 \sin (4 \Theta )+14 \sin (2 \Theta )), \\
    & \mu_{1122} = -\frac{7 \mu_{0}}{2} (3 \cos (2 \Theta )+11), \\
    & \mu_{1133} = -\frac{\mu_{0}}{8} (117 \cos (4 \Theta ) + 107), \\
    & \mu_{1212} = \frac{9 \mu_{0}}{2} (5 \cos (2 \Theta )+9), \\
    & \mu_{1223} = -\frac{ 45 }{ 2 } \mu_{0} \sin (2 \Theta ), \\
    & \mu_{1313} = -\frac{9 \mu_{0}}{8} ( 13 \cos (4 \Theta ) - 29 ), \\
    & \mu_{1322} = \frac{ 21 }{ 2 } \mu_{0} \sin ( 2 \Theta ), \\
    & \mu_{1333} = \frac{3 \mu_{0}}{8} (39 \sin (4 \Theta )-14 \sin (2 \Theta )), \\
    & \mu_{2222} = 77 \mu_{0}, \\
    & \mu_{2233} = \frac{7 \mu_{0}}{2} (3 \cos (2 \Theta )-11), \\
    & \mu_{2323} = -\frac{9 \mu_{0}}{2} ( 5 \cos (2 \Theta ) - 9 ), \\
    & \mu_{3333} = \frac{\mu_{0}}{8} (117 \cos (4 \Theta )-84 \cos (2 \Theta )+415),  
\end{align}
\end{subequations}
where $\mu_{0} = \frac{ 5 }{ 1024 a_d^2 } \sqrt{ \frac{ m }{ \pi \beta_0 } }$. Other non-trivial viscosity terms are specified by the tensor symmetry identities 
\begin{subequations} \label{eq:tensor_symmetries}
\begin{align}
    \mu_{i j m n} &= \mu_{j i m n} = \mu_{j i n m} = \mu_{m n i j}, \\
    \mu_{i j m n} \delta_{i j} &= \mu_{i j m n} \delta_{m n} = \mu_{i j m n} \delta_{i j m n} = 0,
\end{align}
\end{subequations}
where $\delta_{i j m n}$ is 1 if $i=j=k=\ell$ and $0$ otherwise. All other unspecified $\mu_{i j k \ell}$ elements are zero.

\section{ Equilibrium Boltzmann Equation \label{app:Boltzmann_equation} }

In this appendix section, we extend the derivation of the equilibrium Boltzmann Equation found in \cite{Bond65_AW}, to include arbitrary external potentials.

At thermal equilibrium, the left-hand side of the Boltzmann equation is given as
\begin{align}
    \frac{ {\cal D} }{ {\cal D} t } f_0 &= \left( \frac{ \partial }{ \partial t } + v_i \partial_i - \frac{ \partial_i V(\boldsymbol{r}) }{ m } \frac{ \partial }{ \partial u_i } \right) f_0 \\
    &= f_0 \left( \frac{ \partial }{ \partial t } + v_i \partial_i - \frac{ \partial_i V(\boldsymbol{r}) }{ m } \frac{ \partial }{ \partial u_i } \right) \ln f_0,
\end{align}
where $f_0$ is of the form in Eq.~(\ref{eq:equilibrium_ansatz}) and $n_0$ is determined by the form of $V(\boldsymbol{r})$, so that
\begin{align}
    \ln f_0 = \frac{ 3 }{ 2 } \ln\left( \frac{ m }{ 2 \pi } \right) + \ln( n_0 \beta^{3/2} ) - \beta \frac{ m \boldsymbol{u}^2 }{ 2 }. 
\end{align}
The material derivative is defined as
\begin{align}
    \frac{ D }{ D t } = \frac{ \partial }{ \partial t } + U_i \partial_i,
\end{align}
so that the ${\cal D} / {\cal D} t$ operator can be rewritten as
\begin{align}
    \frac{ {\cal D} }{ {\cal D} t } = \frac{ D }{ D t } + u_i \partial_i - \frac{ \partial_i V(\boldsymbol{r}) }{ m } \frac{ \partial }{ \partial u_i }. 
\end{align}
We now treat the derivatives term by term. 

First considering $u_i \partial_i \ln f_0$, we have
\begin{align}
    u_i \partial_i \ln f_0 
    &=
    u_i \partial_i \left[ \ln( n_0 ) + \frac{3}{2} \ln( \beta ) - \beta \frac{ m \boldsymbol{u}^2 }{ 2 } \right], \nonumber\\ 
    &= u_i \partial_i \ln( n_0 ) + \left( \beta \frac{ m \boldsymbol{u}^2 }{ 2 } - \frac{3}{2} \right) u_i \partial_i \ln T \nonumber\\
    &\quad + \beta m u_i u_j \partial_i U_j. 
\end{align}
As for ${ D \ln f_0 / D t }$, we first consider the equations of conservation at thermal equilibrium which read
\begin{subequations}
\begin{align}
    & \frac{ D }{ D t } \ln n_0 = -\partial_j U_j, \\
    & \frac{ D }{ D t } \ln T = -\frac{ 2 }{ 3 } \partial_j U_j, \\
    & \frac{ D }{ D t } U_i = -\frac{ k_B T }{ m } \partial_i \ln( n_0 T ) - \frac{ 1 }{ m } \partial_i V(\boldsymbol{r}).
\end{align}
\end{subequations}
The material derivative of $\ln f_0$ then becomes
\begin{align}
    \frac{ D }{ D t } \ln f_0 
    &= 
    \frac{ D }{ D t } \left[ \ln( n_0 ) + \frac{3}{2} \ln( \beta ) - \beta \frac{ m \boldsymbol{u}^2 }{ 2 } \right] \nonumber\\ 
    &= 
    - \frac{ \beta }{ 3 } m \boldsymbol{u}^2 \partial_j U_j - \beta u_i \partial_i V(\boldsymbol{r}) \nonumber\\
    &\quad - u_i \partial_i \ln( n_0 ) - u_i \partial_i \ln T. 
\end{align}
Finally, the term explicit in the potential is
\begin{align}
    - \frac{ \partial_i V(\boldsymbol{r}) }{ m } \frac{ \partial \ln f_0 }{ \partial u_i } 
    &= 
    \frac{ \beta \partial_i V(\boldsymbol{r}) }{ 2 } \frac{ \partial \boldsymbol{u}^2 }{ \partial u_i } 
    = 
    \beta u_i \partial_i V(\boldsymbol{r}). 
\end{align}
Putting all these terms together, we get
\begin{align}
    \frac{ {\cal D} }{ {\cal D} t } \ln f_0 &= \frac{ D }{ D t } \ln f_0 + u_i \partial_i \ln f_0 - \frac{ \partial_i V(\boldsymbol{r}) }{ m } \frac{ \partial }{ \partial u_i } \ln f_0 \nonumber\\
    &= \bigg[ 
    -\frac{ \beta }{ 3 } m \boldsymbol{u}^2 \partial_j U_j - u_i \partial_i \ln T \nonumber\\
    &\quad\quad  
     - \beta u_i \partial_i V(\boldsymbol{r}) - u_i \partial_i \ln( n_0 )
    \bigg] \nonumber\\
    & + \bigg[ 
    \beta m u_i u_j \partial_i U_j + u_i \partial_i \ln( n_0 ) \nonumber\\
    &\quad\quad + \left( \beta \frac{ m \boldsymbol{u}^2 }{ 2 } - \frac{3}{2} \right) u_i \partial_i \ln T
    \bigg] \nonumber\\
    & + \bigg[ 
    \beta u_i \partial_i V(\boldsymbol{r}) 
    \bigg],
\end{align}
which gives the final result
\begin{align}
    \frac{ {\cal D} }{ {\cal D} t } \ln f_0 
    &=
    \left( \beta \frac{ m \boldsymbol{u}^2 }{ 2 } - \frac{5}{2} \right) u_i \partial_i \ln T \nonumber\\
    &\quad\quad +
    \beta m \left( u_i u_j - \frac{ 1 }{ 3 } \delta_{ij} \boldsymbol{u}^2 \right) \partial_i U_j. 
\end{align} 
This shows that the external potential does not affect the Chapman-Enskog derivation of transport tensors.

\section{ Normal Mode Solutions \label{app:mode_derivation} }

The normal mode solutions of the fluid are obtained by diagonalizing the matrix in Eq.~(\ref{eq:eigensystem}), for which the eigenvectors will be the mode amplitudes, and eigenvalues the associated mode frequencies. To obtain these modes analytically, we first notice that the structure of $\boldsymbol{\Lambda}$ in Eq.~(\ref{eq:transport_rates}) implies that $\xi_y$ is uncoupled from all other variables, leaving us to diagonalize the matrix
\begin{align}
    \boldsymbol{\Omega} 
    &= 
    \begin{pmatrix}
        0 & 0 & c K & 0 \\
        0 & 
        i \Lambda_{11} & i \Lambda_{13} & 0 \\
        \left( \frac{3}{5} + \frac{ \mathscr{E}_{\rm dd} }{ c^2 } \right) c K & 
        i \Lambda_{13} & i \Lambda_{33} & 
        \frac{3}{5} c K \\
        0 & 0 & \frac{2}{3} c K & i \Gamma 
    \end{pmatrix}. 
\end{align}
The frequency solution for $\xi_y$ is then immediately read off as $\omega_{\mu,2} = i \Lambda_{22}$. The remaining solutions are obtained by solving the characteristic polynomial of $\boldsymbol{\Omega}$,
\begin{align}
    \omega^4
    &-
    \omega^3 i ( \Gamma + \Lambda_{11} + \Lambda_{33} ) \nonumber\\
    &-
    \omega^2 \left[ ( \Lambda_{11} \Lambda_{33} - \Lambda_{13}^2 ) + \Gamma ( \Lambda_{11} + \Lambda_{33} ) + c_{\rm dd}^2 K^2 \right] \nonumber\\
    &+
    \omega 
    i \bigg[
    \left( \Lambda_{11} \Lambda_{33} - \Lambda_{13}^2 \right) \Gamma \nonumber\\
    &\quad\quad\quad\quad + 
    \left(
    ( \Gamma + \Lambda_{11} ) c_{\rm dd} - \frac{ 2 }{ 5 } \Gamma c^2
    \right) K^2 \bigg] \nonumber\\
    &+
    \Gamma \Lambda_{11} \left(
    c_{\rm dd}^2 - \frac{2}{5} c^2
    \right) K^2 = 0.
\end{align}
To first order in $\delta$, solutions to the fourth order polynomial above can be computed with Mathematica \cite{Wolfram22}, to give the mode frequencies in Eqs.~(\ref{eq:firstorder_omega_solutions}), and mode vectors in Eqs.~(\ref{eq:propagating_modes}), (\ref{eq:y_shear_mode}), (\ref{eq:x_shear_mode}) and (\ref{eq:density_mode}).

\nocite{*}
\bibliography{main.bib} 

\end{document}